# Generative Adversarial Learning for Intelligent Trust Management in 6G Wireless Networks

Liu Yang, Yun Li, Simon X. Yang, *Senior Member, IEEE*, Yinzhi Lu,

Tan Guo, *Member, IEEE*, and Keping Yu, *Member, IEEE*

*Abstract*—Emerging six generation (6G) is the integration of heterogeneous wireless networks, which can seamlessly support anywhere and anytime networking. But high Quality-of-Trust should be offered by 6G to meet mobile user expectations. Artificial intelligence (AI) is considered as one of the most important components in 6G. Then AI-based trust management is a promising paradigm to provide trusted and reliable services. In this article, a generative adversarial learning-enabled trust management method is presented for 6G wireless networks. Some typical AI-based trust management schemes are first reviewed, and then a potential heterogeneous and intelligent 6G architecture is introduced. Next, the integration of AI and trust management is developed to optimize the intelligence and security. Finally, the presented AI-based trust management method is applied to secure clustering to achieve reliable and real-time communications. Simulation results have demonstrated its excellent performance in guaranteeing network security and service quality.

## I. INTRODUCTION

THE latest fifth generation (5G) technology was originally conceived as the enabler of Internet of Everything (IoE) applications. Currently, 5G is still on the way being deployed for the interconnection between things, people, process, and data in a uniform manner. However, with the unprecedented proliferation of smart terminals and emerging new IoE services, 5G may be unable to satisfy the full demands of the quickly rising technology for future digital society. To meet the upcoming challenges, now it is time to shift the attention toward beyond 5G or the sixth generation (6G) for industry and academia. 5G has taken a significant step toward the redesign of core network, utilization of new frequency bands, and management of advanced spectrum including licensed and unlicensed bands [1]. One key technique, artificial intelligence (AI), is missing in the existing 5G cellular networks, which is widely considered to be one of the most important components for the emerging 6G technique [2]. Through automatic learning and training with big data, AI can drive the interconnection between IoE devices in future 6G wireless networks. The main goals of 6G include not only a greatly improved transportation network, but also a fully autonomous, highly intelligent, and extremely ultra-dense heterogeneous IoE system. To satisfy the requirements of future IoE applications, a space-air-ground-sea integrated wireless communication network will be established for 6G by merging the satellite, aerial network, terrestrial network, and marine network [3].

Because of the dynamic and heterogeneous nature, security is regarded as one of the main concerns in 6G network [4]. However, it is with great challenge to enforce a uniform security system due to decentralized operation and the practical constraints of some network devices. Moreover, AI will be inevitably distributed to the network edge to make the security system more reliable and adaptive. Then the effectiveness and intelligence are key factors that need to be considered when designing the 6G security system. Trust is a potential method that is easy to be implemented and can enhance the confidence of users on introducing autonomic AI-driven security system in 6G. ITU-T has given the definition of trust and keeps working on the designing of trust-based network framework for 6G, and the networking chapter of the first IEEE 6G Summit White Paper has presented the proposal of integrating trust into the 6G wireless networks [5]. To assure the secure end-to-end (E2E) communication, trust has been widely used in ad-hoc networks and social networks. However, how to enhance the intelligence of 6G security by integrating AI into the trust management system and assure reliable E2E communication in IoE system is still a hard work that needs to be further discussed.

Traditional AI techniques have been successfully adopted for trust management, whereas, the lack of training dataset limits the accuracy of trust evaluation in some applications. Due to the dynamic characteristics of 6G wireless networks, it is hard to enrich the training dataset so that traditional AI techniques may not be a good choice. As a promising deep learning method, generative adversarial network (GAN) has recently gained much attention from researchers. The GAN structure consists of a discriminator and a generator, which are trained to identify real samples from fake ones and generate samples according to the learned distribution, respectively [6]. Hence, the potential in learning data distribution and generating the expected data makes GAN very suitable for trust model construction in 6G wireless networks. On the one hand, GAN can be used to synthesize more samples for trust model training. On the other hand, by learning the distribution of the trust information, the commonalities and differences in behaviors of network devices can be analyzed for further trust decision-making.

Based on the above views, the application of GAN on trust management will be discussed in this article. Then intelligent trust management will be studied to guarantee the reliable and real-time communications in 6G wireless networks. The main

Liu Yang, Yinzhi Lu, and Tan Guo are with the School of Communication and Information Engineering, Chongqing University of Posts and Telecommunications; Yun Li (*Corresponding author*) is with the School of Software Engineering, Chongqing University of Posts and Telecommunications; Simon X. Yang is with the Advanced Robotics and Intelligent Systems Laboratory, University of Guelph; Keping Yu is with the Hosei University.

contributions are listed as follows: First, a novel intelligent trust management framework that combines fuzzy logic theory and adversarial learning method is constructed. Then a GAN-based trust decision-making model is proposed, which can be used to determine whether a network device is trustworthy and makes the trust management resilient. Finally, the proposed intelligent trust management method is adopted to guarantee trusted and secure data delivery in heterogeneous 6G wireless networks.

## II. RELATED WORK

Recently, researches have been focused on the issue of AI applications to trust management for secure Internet of Things (IoT) services. Through pattern recognition with AI techniques, malicious IoT devices can be identified and then isolated from the network. In this section, some AI-based trust management methods are first reviewed, and then a potential heterogeneous and intelligent 6G architecture is introduced.

### A. AI applications to trust management

Jayasinghe *et al.* [7] presented a machine learning-enabled trust computational method (MLTC) for reliable IoT services. Individual trust attributes regarding reputation, experience, and knowledge were first computed. Next, principal component analysis (PCA) was used to extract features from the obtained trust attributes. These trust features were labelled based on the algorithm K-Means and finally used to train a support vector machine (SVM) framework for trust classification.

To acquire accurate and robust trust estimation in underwater sensor networks, a synergetic trust model with SVM (STMS) was presented by Han *et al.* [8]. The network was divided into some clusters where the cluster heads and member nodes were synergetic to establish the trust relationship. First, three kinds of trust evidences from the aspects of energy, packet, and communication were gathered by member nodes. Next, the evidences were transmitted to the corresponding cluster heads for further analysis: The algorithm K-Means was first used for labelling, and then a SVM framework was trained to classify trust evidences into various categories. Finally, the constructed trust classification model was returned to member nodes to detect abnormal neighbors.

Jiang *et al.* [9] presented a trust evaluation and update method using C4.5 decision tree (TEUC). First, trust attributes regarding the energy consumption, data consistency, and link failure were collected. After that, a C4.5 decision tree was built based on the normalized and fuzzified trust attributes. Finally, trust classification was performed to get the trust degree, and trust update would be triggered by event or time.

SVM and Dempster-Shafer (DS) theory were combined to present a SVM-DS fusion-based trust management (SDFTM) method by Su *et al.* [10]. Trust evidences from the aspects of packet, data, and energy were first collected. Next, features of these kinds of evidences were separately extracted to further train the corresponding trust classifiers. To obtain the overall trust classifier, DS evidence rules were finally adopted to fuse the trained sub-classifiers.

A fault-tolerant trust mechanism (FTTM) for underwater sensor networks was presented by Han *et al.* [11]. To improve trust evaluation accuracy, an underwater environment model was first constructed where both the mobility and the acoustic communication factors were considered. After that, a trust update method was presented using the environment model and Q-learning, for achieving timely and accurate trust estimation. Finally, to promote resource utilization, a trust redemption model was presented to provide additional opportunities of rejoining the network for the normal sensor nodes considered to be malicious due to random network failures.

To achieve the secure IoT services, a multi-perspective trust management model (MPTM) was proposed by Bahutair *et al.* [12]. First, trust attributes were obtained from the perspectives of service, device, and owner according to the prepared training dataset. Next, the influence degrees of attributes on trust were estimated and further adopted to train a neural network-driven trust model. Finally, the attributes of an IoT service from all perspectives acted as the inputs of the trained neural network to get the output trust grade.

Liu *et al.* [13] proposed an efficient detection mechanism to avoid conditional packets manipulation attacks in IoT network. To evaluate the trust value of each IoT device, a regression model was first trained based on the reputation of the related routing paths. Then, trust values were computed and grouped into three clusters corresponding different trust levels. The trust grade of any IoT device was finally acquired by determining the cluster to which the trust value of the corresponding IoT device belongs.

In view of the above AI applications to trust management, trust evaluation usually begins with trust evidence collection. If a certain number of evidences are collected, feature extraction can be performed to reduce the data dimension. Next, trust evidences are labelled with clustering algorithms and then used to train the trust classification model, which may be constructed with classification algorithms like regression analysis, decision tree, SVM, and neural network. Finally, trust decision-making is performed based on the trained trust model while the model update can be carried out if some newest trust information is available. Although many traditional AI algorithms have been adopted for trust model construction, the performances of the corresponding systems strongly depend on the training dataset scale. Then how to use AI techniques for trust management and improve the trust evaluation accuracy in 6G wireless networks is a great challenge to be overcome.

### B. Architecture for 6G

Heterogeneity and intelligence integrated wireless networks can be considered as a potential 6G architecture. To achieve global seamless coverage, the space-air-ground-sea integrated 6G wireless networks can promote the interconnection between heterogeneous network devices in IoE applications. Then from the perspective of heterogeneity, the 6G integrated networks can be distributed into four tiers [3]: The first tier is the marine network, followed by the terrestrial network and aerial network. The last tier is the satellite network. To fully achieve the 6G goals, the distributed heterogeneous networks require large numbers of AI services from the initial data sensing to the final smart applications. Then from the perspective of intelligence, four layers are included in the 6G architecture [2]: The bottom is the intelligent sensing layer, followed by the intelligent edge and intelligent control layers. The top layer is smart application.

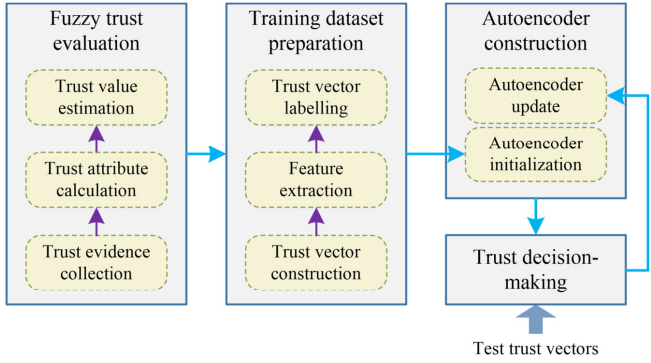

FIGURE 1. Framework of GITM.

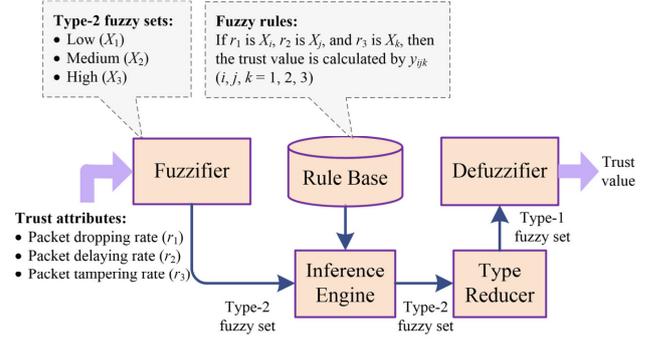

FIGURE 2. Fuzzy trust evaluation model.

Based on this heterogeneous and intelligent 6G architecture, a distributed AI-driven trust management mechanism would be a better approach. It enables each smart device to organize its own knowledge base autonomously for pattern recognition and trust classification against the attacks. Some smart devices with more storage capacity and higher computing ability can even perform trust prediction via AI techniques and big data. With the interaction between smart devices, the knowledge base continues expanding while the trust decision-making accuracy can be continuously improved.

## III. GENERATIVE ADVERSARIAL LEARNING-ENABLED INTELLIGENT TRUST MANAGEMENT

Due to the excellent performance in deep learning, GAN is introduced for trust management. Then a generative adversarial learning-enabled intelligent trust management (GITM) method is presented. The framework of GITM is given in Fig. 1, and the main courses include the fuzzy trust evaluation, training dataset preparation, GAN-based autoencoder construction, and trust decision-making. Details will be discussed below.

### A. Fuzzy trust evaluation

To guarantee the reliability and timeliness of data delivery in 6G wireless networks, the data-plane information regarding network change when attacks happen needs to be gathered for trust evaluation. Then the trust evidences to be collected should include dropping, delaying, tampering, and timely forwarding behaviors. If the behavior of a network device is overheard in one data transmission round, trust attributes including packet dropping, delaying, and tampering rates can be first calculated based on a certain number of behaviors of this device overheard in recent rounds. After that, the current trust value of this device can be estimated according to the trust attributes.

Due to the extremely dynamic and heterogeneous nature of 6G wireless networks, that device faults and link failures may happen inevitably makes it difficult to determine whether an overheard adverse behavior comes from a malicious attack. Hence, trust uncertainty problem exists that may reduce the trust evaluation accuracy. Fuzzy logic is an effective theory to deal with such problem, and then a type-2 fuzzy logic model is introduced for trust evaluation due to its better performance than the type-1 counterpart [14]. As shown in Fig. 2, the fuzzy trust evaluation model consists of the fuzzifier, inference engine, type reducer, and defuzzifier modules. To acquire the output trust value of a network device, the related trust attributes are used as the inputs of the trust evaluation model. Fuzzification operation is firstly performed to convert the input trust attributes into the type-2 fuzzy sets low, medium, and high. Each type-2 fuzzy set has an upper membership function and a lower one whose shape can be Gaussian and piecewise linear. Then based on the rules that define the input-output mappings, type-2 output fuzzy sets can be obtained via fuzzy inferring. An example of fuzzy rule is like this: If dropping rate is low, delaying rate is low, and tampering rate is low, then the output trust value is calculated by a formula. The formulas designed for the output trust values should reflect the difference in trust levels between different fuzzy rules. Next, type reduction is carried out to get the type-1 output fuzzy set, and the final trust value can be calculated via defuzzification.

### B. Training dataset preparation

Due to the absence of prior trust decision-making standard in 6G wireless networks, it is hard to determine whether a network device is trustworthy or not based on the estimated trust value. Then an intelligent trust classification model is necessary for automatic pattern recognition and trust decision-making. To improve the decision accuracy, the trust vector consisting of several consecutive trust evaluation results should be used to test the corresponding network device. Hence, a dataset that contains a certain number of trust vectors needs to be prepared in advance to train the trust model.

As the network operates, trust vectors can be acquired during the interactions between network devices. Whereas, these trust vectors should be labelled before the trust model training. Then to prepare the dataset, feature extraction is firstly performed for all the obtained trust vectors to reduce the data dimension. The algorithm PCA can be selected to achieve this purpose by extracting the main data features. Next, the algorithm K-Means is adopted to cluster all feature vectors into three trust groups that correspond to the higher, medium, and lower trust levels. After that, each trust vector can be labelled by determining the group to which its feature vector belongs based on the distances to the means of groups. The obtained trust vectors labelled with the higher and medium trust levels are finally used to build the training dataset, while the corresponding network devices are considered trustworthy and uncertain, respectively.

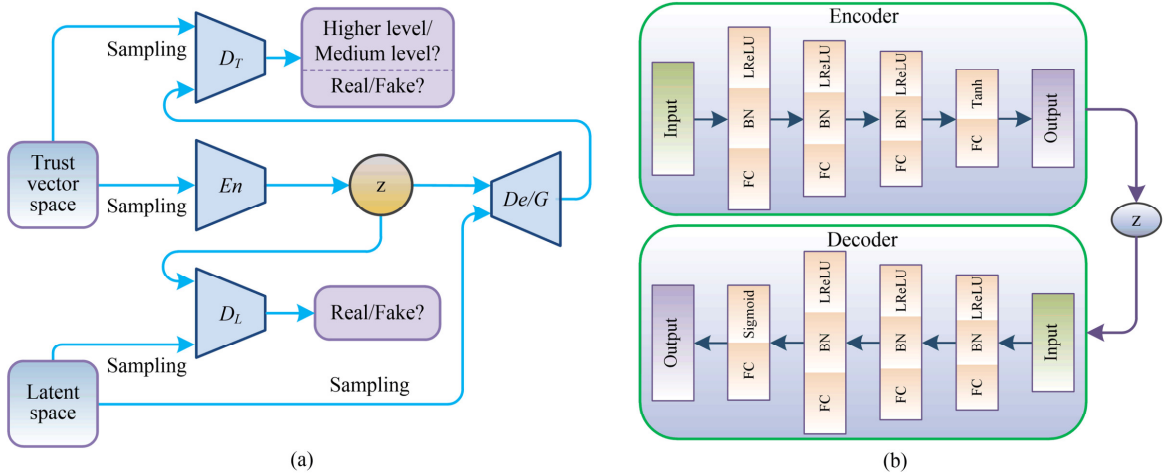

FIGURE 3. The autoencoder: (a) Autoencoder architecture; (b) Detailed network structure of the autoencoder.

*C. GAN-based autoencoder construction*

Whether a network device is trustworthy can be determined via feature extraction and labelling for its trust vector, whereas, good performances require a large number of trust vectors to be used together for clustering. Inspired by the generative model of GAN, the problem of trust decision-making can be solved by estimating whether the test trust vectors satisfy the distribution of those labelled with a higher trust level. Then an autoencoder architecture with pair of GANs are constructed, as shown in Fig. 3. Four modules are included that are the trust discriminator ($D_T$), decoder/generator ($De/G$), latent discriminator ($D_L$), and encoder ($En$). Both the encoder and decoder consist of one input layer, one output layer, and four fully connected layers.

As the core components of the autoencoder architecture, the encoder and decoder learn the distributions of latent data and trust vectors respectively, for the purpose of synthesizing the data according to these distributions. The trust discriminator and latent discriminator are trained to distinguish the real data from the fake one with respect to the trust vector space and latent space respectively. Moreover, the trust discriminator is trained to determine whether a network device is trustworthy or uncertain by classifying its trust vector into either the group with a higher trust level or the one with a medium level. For both the encoder and decoder, in the first three fully connected layers, *batch normalization* is utilized while *Leaky Relu* is used for activation on the output. In addition, *Tanh* and *Sigmoid* are used as the activation in the last fully connected layers of the encoder and decoder respectively. During the training process, a batch of trust vectors is as the input of the encoder in each epoch while the method *Adam* is adopted for optimization. For both the encoder and decoder, score losses are firstly computed with the *least square* method according to the expected scores and those given by the discriminators and then used to update the parameters. In addition, a reconstruction loss defined by the mean absolute difference between the input and output of the autoencoder is also fed back to the decoder for update.

Due to the dynamic nature of 6G wireless networks, the initially trained autoencoder may not always be effective in the future network operation phase. Since the autoencoder is finally used to detect whether the test trust vectors satisfy the training data distribution, some latest test trust vectors can be selected to retrain the autoencoder. Whereas, to avoid the drift of network parameters, only those test trust vectors that strictly satisfy the training data distribution can be used for further autoencoder update. If several batches of eligible trust vectors are collected, the training dataset will be updated, and then the autoencoder retraining process can be activated.

*D. Trust decision-making*

To determine whether a network device is trustworthy or not, the newest trust vector of this device needs to be acquired firstly. Next, the trust vector is used as the input of the autoencoder to test whether it satisfies the distribution of the training data. If the output deviates from the input, then this device is directly considered to be untrusted. Otherwise, the trust discriminator is used to classify the trust vector into either the group with a higher trust level or the one with a medium level, and the device can either be trustworthy or uncertain.

Due to the occasional faults or link failures in dynamic 6G wireless networks, a normal network device may accidentally be detected to be uncertain. To further determine whether such device is trustworthy or not, trust recommendation should be introduced for synergetic detection. If several trusted neighbors have fortunately recommended the uncertain device, then this device can temporarily be considered as the trustworthy.

*E. Implementation of the trust management method*

For improving the overall security performance of the 6G wireless networks, all network devices should collaborate with each other to maintain and share the trust information. Usually, different kinds of network devices are with different computing and storage capabilities in the heterogeneous network, and they can be classified into the super, advanced, and generic groups with respect to their capabilities. For the implementation of the intelligent trust management method, the three kinds of devices are burdened with different tasks. The super network devices need to build the training dataset and then construct the initial autoencoder architecture in the initial network operation phase. In addition, each super device should share parameters of the well-trained autoencoder with the neighbor advanced devices. As the network operates, both the super and advanced devices periodically perform trust decision-making based on their own autoencoders and the inferred trust vectors, while each one should independently select some eligible test trust vectors for further autoencoder update. As for the generic network devices, they cannot construct or update an autoencoder due to limited computing and storage capabilities. Then they perform trust decision-making according to the evaluated trust values and the thresholds recommended by their super or advanced neighbors.

The procedure of the proposed intelligent trust management method GITM is given as follows: First, through the courses of fuzzy trust evaluation, training dataset preparation, and model training, super devices construct the initial autoencoders and then share parameters with neighbor advanced devices. Next, super and advanced devices periodically perform independent trust decision-making and autoencoder update. Once a device has completed the update, it sorts all trust values of the vectors labelled with a higher trust level in its current training dataset in descending order and removes the last 25% of values. Then the minimum of the rest values is considered as the trust threshold and recommended to its neighbor generic devices. Finally, each generic device eliminates the received abnormal thresholds via clustering, and then takes the average of the remaining ones as its own trust threshold for trust decision-making.

## IV. APPLICATION OF PROPOSED TRUST MANAGEMENT METHOD TO SECURE CLUSTERING

To guarantee the reliability and timeliness of data delivery in 6G wireless networks, the proposed trust management method is applied to data routing. Then a trusted and secure clustering protocol is presented.

To group secure clusters, the heads need to be selected firstly. Then any non-cluster head device can join a cluster in which the head is trustworthy. In a certain round $r$, a device becomes the head according to a threshold $Th_{CH}$ [14] if it is eligible in the current round. If a device successfully becomes a cluster head, it informs all neighbors its election message. To assure reliable and secure transmission, each device is better to avoid joining the cluster where the head is detected to be malicious. For any non-cluster head device, it takes all neighbor heads as its own candidates and desires to find the nearest trusted one to join the corresponding cluster. Any super or advanced device performs the trust decision-making to decide whether a candidate can be trusted based on its own autoencoder and the trust vector of the candidate. While a generic device selects the trusted candidate according to the trust values of all candidates and a threshold acquired based on the recommendations from the neighbor super or advanced devices. A non-cluster head device finally has to act as the head if it does not have a proper head while being eligible in the current round.

If cluster formation has been done, a member device begins to transmit data to the cluster head within an allocated time slot and then keeps monitoring the followed transmission behaviors for evidence collection. Based on the newest evidences, the current trust value of the cluster head is computed while the trust vector can be updated. After trust decision-making, the eligible trust vector is reserved for the next autoencoder update.

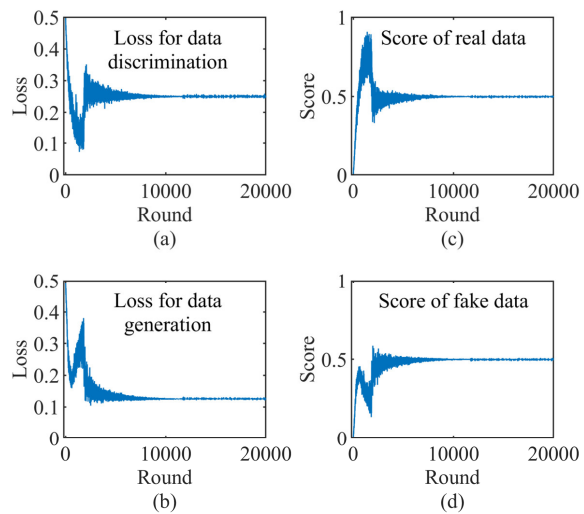

FIGURE 4. The losses and scores of GAN in each training round.

## V. PERFORMANCE EVALUATION AND SECURITY DISCUSS

### A. Experiment settings

For performance evaluation, a 6G network with 100 smart devices is considered, where three kinds of malicious devices with higher, medium, and lower attack capabilities are included and account for 30%, 40%, and 30% of the total number of the malicious. In addition, the percentages of the super, advanced, and generic network devices are 30, 50, and 20. Another two trust models called TECC [15] and TEUC are introduced for performance comparison, since the cloud model is adopted for fuzzy trust evaluation in TECC, and C4.5 trust decision tree is constructed in TEUC to make the trust management intelligent. The initial energy for secure data delivery is 1.3J, while other energy parameters are with the same settings as those in [14].

### B. Simulation results

The losses for data discrimination and generation, and the scores for real and generated fake data during the GAN training process are given in Fig. 4. It verifies that the discriminator and generator are two adversarial game players, if one gains, then another loses. Therefore, the loss and score fluctuate and finally tend to be stable, while the game achieves a relative equilibrium state. Then, the generator can generate the data that satisfies the learned distribution, while the discriminator discriminates the real data from the generated fake one just by guessing.

The comparison of security performance including security guarantee rate and total number of attacks is shown in Fig. 5. Here security guarantee rate is defined as the ratio of the total number of rounds from when attacks no longer happen until the end of the network operation. The results show that the security guarantee rate of the network using GITM is the highest, while that of the network using TECC keeps the lowest value 0. Especially, the security guarantee rates are above 39% and 16% for a network with 50% malicious devices if GITM and TEUC are adopted respectively. This case indicates that it is possible for the network using GITM or TEUC to avoid the attacks if enough trust evidences are available. The results also show that, the network using GITM has obviously fewer malicious attacks than that using TEUC or TECC if the percentage of malicious devices is bigger than 20. Hence, GITM outperforms other two methods if the percentage of malicious devices is significant. Whereas, the use of synergetic detection in GITM also brings the risk of attacks, so that GITM does not outperform other two methods in protecting the network from malicious attacks if the percentage of the malicious is small.

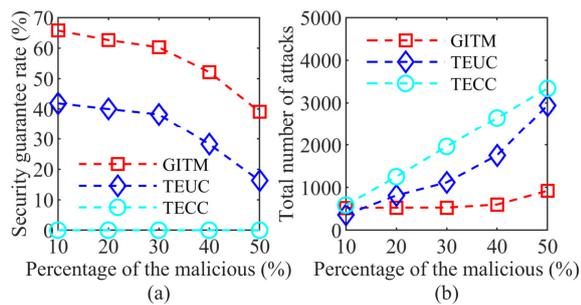

FIGURE 5. Security performance: (a) Security guarantee rate; (b) Total number of attacks.

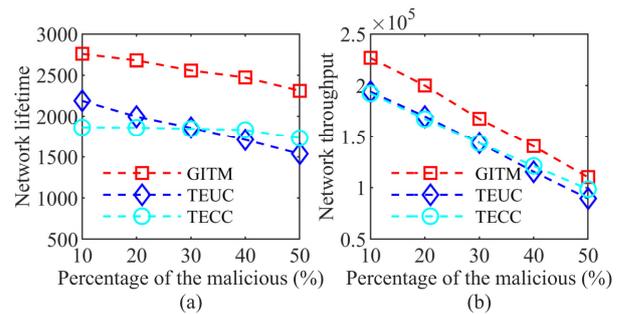

FIGURE 6. Network performance: (a) Network lifetime; (b) Network throughput.

The comparison of network performance including lifetime and throughput are given in Fig. 6. Results show that both the network lifetime and throughput decrease when the percentage of malicious devices increases. The reason for this case is that more malicious devices are detected and then isolated from the network if the percentage is bigger, so that heavier data tasks are left for most of the remainder. Owing to the synergetic detection adopted in GITM, the resilience of trust management can be improved. Then the devices that occasionally behave maliciously due to faults or link failures still have the chance to be trusted, so that they can keep working in the network to share tasks and contribute the throughput. Hence, experiment results also show that the network using GITM has better performance in lifetime and throughput than that with TEUC or TECC, and the network lifetime and throughput are increased by more than 30% and 10% respectively if the percentage of the malicious is 50.

## VI. CONCLUSION AND FUTURE WORK

To enhance the intelligence and security of 6G wireless networks, AI application to trust management is explored in this article. First, type-2 fuzzy logic is adopted to evaluate the trust of devices while alleviating the trust uncertainty problem. Then, through feature extraction and data labelling for the built trust vectors, a dataset can be prepared to train the GAN-based autoencoder. Finally, trust decision-making can be carried out according to the trained autoencoder and the test trust vectors, while a synergetic detection scheme and an autoencoder update method are presented to enhance the resilience and timeliness of trust management. The proposed trust management method considers both the heterogeneity and intelligence requirements of 6G wireless networks, and it is an attempt to address the 6G trust problem by combining fuzzy logic theory and adversarial learning method. Then more studies on 6G trust management with AI techniques may be inspired.

In future work, trust management in 6G wireless networks will be further studied from the following three aspects: First, to adapt to the dynamic and heterogeneous nature of the network, fast convergent algorithms will be focused on rapid trust model training. The combination of semi-supervised learning methods and GAN is a potential direction to address this issue. Second, to improve the network resource utilization, trust redemption methods will be researched to incentivize the malicious devices to provide active services. Contextual GAN may be an effective technique due to its potential in behavior prediction. Finally, to enable the new network devices to quickly establish the trust relationship with others, reliable trust information maintaining and sharing mechanism will be studied. The immutability of blockchain makes it a potential solution to achieve this purpose.

ACKNOWLEDGMENT

This work was supported by the Science and Technology Research Program of Chongqing Municipal Education Commission under Grant KJQN202000641.

BIOGRAPHIES

**Liu Yang** (yangliu@cqupt.edu.cn) received his B.S. degree in Electronic Information Science and Technology from Qingdao University of Technology, Shandong, China, in 2010, and Ph.D. degree in Communication and Information Systems at the College of Communication Engineering, Chongqing University, Chongqing, China, in 2016. He is now a lecturer in Chongqing University of Posts and Telecommunications, Chongqing, China. His research interests include Internet of Things, data analysis, and artificial intelligence.

**Yun Li** (liyun@cqupt.edu.cn) is currently a professor with the School of Software Engineering, Chongqing University of Posts and Telecommunication, Chongqing, China. He received his Ph.D degree in communication engineering from the University of Electronic Science and Technology of China. His research interests include mobile cloud/edge computing, cooperative/relay communications, green wireless communications, wireless ad hoc networks, sensor networks, and virtual wireless networks. He is the Executive Associate Editor of Elsevier/CQUPT Digital Communications and Networks (DCN). He is on the editorial boards of *IEEE Access* and *Security and Communication Networks*.

**Simon X. Yang** [S'97–M'99–SM'08] (syang@uoguelph.ca) received the B.Sc. degree in engineering physics from Beijing University, Beijing, China, in 1987, the first of two M.Sc. degrees in biophysics from the Chinese Academy of Sciences, Beijing, China, in 1990, the second M.Sc. degree in electrical engineering from the University of Houston, Houston, TX, in 1996, and the Ph.D. degree in electrical and computer engineering from the University of Alberta, Edmonton, AB, Canada, in 1999. Currently, he is a Professor at the University of Guelph, Guelph, ON, Canada. His research interests include robotics, intelligent systems, wireless sensor networks, control systems, and computational neuroscience.

**Yinzhi Lu** (henanluyinzhi@163.com) received her M.S. degree in Communication and Information Systems from Chongqing University, Chongqing, China, in 2014. She is currently pursuing the Ph. D. degree in Information and Communication Engineering with the School of Communication and Information Engineering, Chongqing University of Posts and Telecommunications, Chongqing, China. She was a teaching assistant with the School of Electronic Information Engineering, Yangtze Normal University from 2014 to 2019. Her current research interests include Internet of Things, time sensitive network, and artificial intelligence.

**Tan Guo** [S'17–M'22] (guot@cqupt.edu.cn) received his M.S. degree in Signal and Information Processing from Chongqing University, Chongqing, China, in 2014, and Ph.D. degree in Communication and Information Systems from Chongqing University, Chongqing, China, in 2017. He is now a lecturer in Chongqing University of Posts and Telecommunications, Chongqing, China. His research interests include Internet of Things, biometrics, pattern recognition, and machine learning.

**Keping Yu** [S'11–M'17] (keping.yu@ieee.org) received the M.E. and Ph.D. degrees from Waseda University, Tokyo, Japan, in 2012 and 2016, respectively. He is currently an associate professor at Hosei University and a visiting scientist at the RIKEN Center for Advanced Intelligence Project, Japan. His research interests include smart grids, information-centric networking, The Internet of Things, artificial intelligence, blockchain, and information security.